\documentclass[pdflatex,sn-mathphys-num]{sn-jnl}

\usepackage{graphicx}%
\usepackage{multirow}%
\usepackage{amsmath,amssymb,amsfonts}%
\usepackage{amsthm}%
\usepackage{mathrsfs}%
\usepackage[title]{appendix}%
\usepackage{xcolor}%
\usepackage{textcomp}%
\usepackage{manyfoot}%
\usepackage{booktabs}%
\usepackage{algorithm}%
\usepackage{algorithmicx}%
\usepackage{algpseudocode}%
\usepackage{listings}%
\usepackage{bm} 
\usepackage{blindtext,alltt}
\usepackage{multirow}
\usepackage{bm} 
\newcommand{\hodge}{{\star}}
\usepackage{graphicx}
\usepackage{breqn}
\usepackage{tikz-cd}
\DeclareMathOperator{\Tr}{Tr}

\theoremstyle{thmstyleone}%
\newtheorem{theorem}{Theorem}
\theoremstyle{thmstyletwo}%

\theoremstyle{thmstylethree}%

\begin{document}

\title[Connecting 2-Forms, Conformal Transformations, Curvature Invariants and Topological Classes in Einstein Spacetimes]{Connecting 2-Forms, Conformal Transformations, Curvature Invariants and Topological Classes in Einstein Spacetimes}


\author*[1]{\fnm{Jack C. M.} \sur{Hughes}}\email{100060536@ku.ac.ae}

\author[1,2]{\fnm{Fedor V.} \sur{Kusmartsev}}\email{fedor.kusmartsev@ku.ac.ae}

\affil*[1]{\orgdiv{College of Art and Science}, \orgname{Khalifa University}, \orgaddress{\city{Abu Dhabi}, \postcode{PO Box 127788}, \country{UAE}}}

\affil[2]{\orgname{Landau Insititute for Theoretical Physics}, \orgaddress{\city{Moskva}, \postcode{119334}, \country{Russia}}}


\abstract{The unique Nature of the Lorentz group in four dimensions is the root cause of the many remarkable properties of the Einstein spacetimes, in particular their operational structure on the 2-forms. We show how this operational structure can be used for two ends. First, it allows for a simple generalization of the Birkhoff theorem to Schwarzschild (A)de-Sitter spacetime. Second, it provides the means to construct an Abelian endomorphism group on the space of 2-forms. It is observed that taking the trace over this group element-wise induces a further Abelian group which may be identified with a tensor representation of conformal transformations, giving Einstein spacetimes access to their own conformal equivalence class. A further trace over the group yields the curvature invariants of the spacetime. The Kretschmann scalar becomes the topological Euler density, which may be linked in a simple way to the Hawking temperature of horizons.}

\keywords{General Relativity, Einstein Spacetime}



\maketitle

\section{Introduction}\label{sec1}

Throughout its development, General Relativity (GR) has been examined from various perspectives \cite{1,2}. One of its most notable attributes is its connection to modern mathematics, particularly through its association with smooth manifolds. This intersection has yielded a number of significant results: the topological theorems regarding singularities and horizon kinematics \cite{3,4}, global existence and stability analyses \cite{5,6}, algebraic decomposition and categorization schemes \cite{7,8, a_1}, and more recently, the identification and interrelations between the numerous alternative formulations of GR itself \cite{9}. As a theory governing the causal structure of spacetime, these findings have profound implications across the spectrum of modern physics, particularly in the realm of quantum gravity.

While these results and theorems can be broadly applicable when considering the topology of spacetime, it is crucial to acknowledge that once specific dimensions and signatures are defined, there will be results specific to these particular cases. There are two primary reasons for this: (i) The choice of dimension dictates the degrees of freedom of the curvature tensor \cite{10} and (ii) The smooth differential operations (for instance the Hodge dual) are influenced by these choices \cite{19, e_1}. This observation often serves as a guiding principle for understanding the fundamental properties and formulation of the field equations, and subsequently, for exploring alternatives to GR \cite{9, e_2, e_3, e_4, e_5}.

An important example of this phenomenon is found in the statement of the Lovelock theorem: the unique choice for a 2-tensor constructed from the metric and its first and second derivatives, having vanishing divergence, is the Einstein tensor together with $\Lambda g_{\mu\nu}$ \cite{11}. Lovelock was able to prove this in four-dimensions without relying on linearity in the second derivatives of the metric, and then later without assuming symmetry in indices \cite{12}. As the dimensionality of spacetime increases, additional terms satisfying these assumptions become possible, and one loses the unique nature of the field equations \cite{13}.

The investigation of four-dimensional spacetime is a cornerstone of modern physics and one which has revealed the unique geometric characteristics of the Lorentz group. Only in four dimensions is the Riemann tensors basis in the product space of 2-forms decomposable \cite{9, 14, 15}. The reason for this lies in the operation of the Hodge dual on 2-forms, which (for Lorentz signature) assumes complex eigenvalues. This allows precise operational statements regarding the structural properties of spacetime to be formulated within the domain of 2-forms. A notable result is that a spacetime is Einstein ($R_{\mu \nu} = \Lambda g_{\mu \nu}$) if and only if the Riemann tensor commutes with the Hodge dual. This insight, expounded upon in Section 2, culminates in a straightforward generalization of the Birkhoff theorem \cite{16} to asymptotically (A)de Sitter spacetimes.

Furthermore the Weyl tensor assumes a critical role in both the 2-form and Lorentz decomposition of the Riemann curvature, where four dimensions is the minimum number required such that it does not vanish identically \cite{10, 17}. For all dimensions greater than or equal to four, this vanishing then occurs if and only if spacetime is locally conformally flat by the Weyl-Shouten theorem \cite{7, 18}. The conformal transformations - those mappings that enlarge the diffeomorphisms to include only angle preservation - emerge as indispensable tools within GR, bridging a gap between local and global properties of spacetime \cite{4, 7, 17, 18, 19, 20, 21}. One such illustration is the conformal equivalence theorem, which demonstrates any analytic $f(R)$ theory is equivalent to (by a conformal transformation) GR minimally coupled to a self-interacting scalar field \cite{22}. Consequently, higher-order gravity theories introduce additional degrees of freedom, allowing them to be constrained \cite{23, 24}. Beyond General Relativity, the influence of the conformal group extends to a wide array of physically significant scenarios. Conformal methodologies have proven crucial in condensed matter and statistical mechanics, particularly in computations involving correlation functions and the theory of phase transitions \cite{25, 26, 27}. The critical collapse analysis of black holes offers an interesting bridge between the phase transitions and GR in this sense, since a universal critical exponent is observed \cite{28}.  This interplay between geometry, symmetry, and conformal transformations opens new avenues for understanding the fundamental nature of spacetime, as we have seen for example in the case of the conformal field theories \cite{29}.

Once again, in four dimensions these properties manifest themselves in important ways. There are a handful of objects that remain invariant under the conformal transformations in GR, the Weyl tensor being one such example that is independent of the dimensionality of spacetime \cite{17, 18}. In addition, those energy-momentum tensors for which the trace is vanishing prove to be conformally invariant. Curiously, the pure Maxwell field satisfies this constraint in four dimensions \cite{19}. Therefore, the light cone determines both the causal and conformal properties of spacetime simultaneously.  Closely connected is the conformal invariance of the Hodge dual acting on 2-forms (or more generally middle degree forms in even dimensions) \cite{9, 19}. It is possible to show that knowledge of the Hodge dual on 2-forms determines the conformal metric, another result unique to four dimensions \cite{9, 30}. Exactly what role the conformal structure plays at the foundational (or perhaps axiomatic) level in GR is far from clear at present, although these points and their formal development have a direct connection to the EPS axiomization scheme \cite{31}, since there are necessarily relations formed between the projective and conformal structures of spacetime \cite{32}. This relationship is something we see quite explicitly in the special case of de Sitter spacetime. de Sitter holds a fundamental position in General Relativity, serving as the dynamical ground state known prior to the solving of the field equations. It is understood as a quotient space, with local kinematics governed by the de-Sitter group which encompasses proper conformal transformations, sourced by an invariant length scale determined by the non-vanishing of the cosmological constant \cite{33}.
 
The arguments above suggest that the four-dimensional Einstein spacetimes (viewed as an extension of de Sitter) offer a particularly effective framework to explore the relationship between differential and conformal structure. Here one has both the Hodge dual, which maintains its form under conformal transformations, and the Riemann tensor, which does not. The essential element is that knowledge of the Hodge dual as an operation on the 2-forms determines the conformal metric. Since in the four-dimensional Einstein spacetimes these operations commute, they must share spectral data. It is not unreasonable to suppose this culminates in the Riemann tensor possessing knowledge of the conformal metric. The objective of this study is to affirm this expectation and draw attention to intriguing mathematical developments occurring within the conformal class of four-dimensional Einstein spacetimes. In section 3, we demonstrate the feasibility of constructing a group of endomorphisms that act on the space of 2-forms through the identification of an appropriate "inverse" Riemann tensor in an operational sense. In the limit where the commutator of the Hodge dual and the Riemann tensor approaches zero (or when the spacetime is Einstein), this group exhibits an Abelian structure. Presently, it remains uncertain what this group operation exactly preserves in terms of action on the 2-forms.

With the Einstein condition in effect, the tensor trace over this group results in a well-defined element-wise reduction of the group. In section 4, we show how this leads to a distinct Abelian group that can be associated with conformal transformations: due to a theorem of Petrov \cite{a_1}, such a trace is always proportional to the Kronecker-delta for any element of the endomorphism group. Acting then on the metric results in a in an overall functional multiplication, which is precisely a conformal transformation. We contend then that the unique attributes of four-dimensional Einstein spacetimes endow them with an understanding of their own conformal equivalence class $[g]$, which is encoded within the Riemann tensor. Notably, the cosmological constant assumes an active role as the generator of homotheties, representing scaling transformations within the conformal class.

In addition to this, one can take the double trace and produce scalar functions from the group of 2-form endomorphisms. We consider briefly the result of this in the final section. Such functions are interesting in themselves, since they - or at least some subset thereof - may be identified with curvature invariants and the integrands of topological invariants. The Einstein spacetimes have the further unique property that their Euler-characteristic is determined completely by the Krestchmann scalar (by the Gauss-Bonet theorem \cite{b_1, b_2}), which is equivalent to demanding that $G$ be Abelian. We end with comments on the potential role of this observation in the dynamics of spacetime.

\section{2-Forms in Four Dimensions}
Within GR one considers spacetime as a differential manifold $\mathcal{M}$ whose structure is specified by both the metric compatibility and vanishing torsion conditions. As a consequence of this, the Riemann tensor possesses a number of symmetries \cite{10, 17}. Regardless of the dimensionality of $\mathcal{M}$, these properties are such that it may be viewed as an operator on the 2-forms - more specifically an endomorphism - under index contraction:
\begin{equation}
    \omega_{\mu \nu} = R^{\alpha \beta}_{\mu \nu} \omega_{\alpha \beta}, \quad \omega \in \Lambda^2(\mathcal{M}),
\end{equation}
where $\Lambda^2(\mathcal{M})$ denotes the space of 2-forms on $\mathcal{M}$. This property is of course well known: the mapping has the same structure in the space of bi-vectors and forms the conceptual basis of the Petrov classification \cite{7, a_1, a_3}. Indeed, since $\mathcal{M}$ is equipped with a metric the spaces of 2-forms and bi-vectors are isomorphic, which will become important later (see section 5.1). Since the Riemann tensor is antisymmetric in both pairs of indices, but symmetric over the pair exchange, it actually constitutes a symmetric matrix over the product space of 2-forms \cite{9, 15}. Specializing now to four-dimensions, this representation exhibits additional structure. The reason is that the Hodge dual is now also an operator on 2-forms, with the additional property that it square to minus the identity,
\begin{equation}\label{eq:2}
     \big(\hodge\big)^2 = \frac{1}{4} \varepsilon_{\mu \nu}^{\alpha \beta} \varepsilon_{\alpha \beta}^{\rho \sigma} = - \frac{1}{2} \big(\delta^{\rho}_{\mu}\delta^{\sigma}_{\nu} - \delta^{\rho}_{\nu}\delta^{\sigma}_{\mu}\big) \equiv  -\mathbb{I}_{\Lambda^2(\mathcal{M})}.
\end{equation}
Therefore, the space of 2-forms is decomposed into a direct sum of two subspaces corresponding to the two possible eigenvalues of the Hodge dual. The Riemann tensor subsequently decomposes but now over the product space, giving (thanks to symmetry) three distinct elements: self-dual self-dual, anti-self-dual anti-self-dual, and mixed. Now, for all dimensions $n \geq 4$ the Riemann tensor also admits its irreducible Lorentz decomposition \cite{10, 35, 36}
\begin{equation}\label{eq:3}
     R_{\mu \nu \rho \sigma} = C_{\mu \nu \rho \sigma} + \frac{1}{2}\big(g_{\mu \rho} R_{\nu \sigma} - g_{\mu \sigma} R_{\nu \rho} - g_{\nu \rho} R_{\mu \sigma} + g_{\nu \sigma} R_{\mu \rho}\big) - \frac{R}{6}\big(g_{\mu\rho} g_{\nu \sigma} - g_{\mu \sigma} g_{\nu \rho}\big).
\end{equation}
The point here is that these two decompositions cannot be independent. A key result unique to four-dimensions is that a spacetime is Einstein if and only if the mixed component of the Riemann tensor vanishes, which is equivalent to the statement that it commutes with the Hodge dual as operations on 2-forms \cite{9, 15},
\begin{equation}\label{eq:4}
    [\text{Riemann} , \hodge] = 0 \quad \Leftrightarrow \quad R_{\mu \nu} = \Lambda g_{\mu \nu},
\end{equation}
which is a corollary of Petrov's classification.

\subsection{Birkhoff Theorem for Einstein spacetimes}

The importance of this result lies in the fact that it is purely operational: the commutator defined above will hold for any choice of 2-form in $\Lambda^2(\mathcal{M})$. The power of this can be exhibited for example in the following statement:

\begin{theorem}\label{thm1}
Any spherically symmetric four-dimensional Einstein spacetime is necessarily Schwarzschild-(A)de Sitter.
\end{theorem}

Previously, this has been shown in a manner analogous to the Birkhoff theorem \cite{16, 34}. However, one can now show this without appealing to the field equation via the operational character of (\ref{eq:4}). Consider an arbitrary 2-form on spacetime; this is defined up to 6 arbitrary functions of the spacetime coordinates. Suppose we have 
\begin{equation}
    \omega = \omega_{\alpha \beta} dx^{\alpha} \wedge dx^{\beta}
\end{equation}
with
\begin{equation}\label{eq:6}
    \omega_{\alpha \beta} =  \begin{pmatrix}
    0 & -A(t,r, \theta, \phi) & -B(t,r, \theta, \phi) & -D(t,r, \theta, \phi) \\
    A(t,r, \theta, \phi) & 0 & -C(t,r, \theta, \phi) & -E(t,r, \theta, \phi) \\
    B(t,r, \theta, \phi) & C(t,r, \theta, \phi) & 0 & -F(t,r, \theta, \phi) \\ 
    D(t,r, \theta, \phi) & E(t,r, \theta, \phi) & F(t,r, \theta, \phi) & 0 \end{pmatrix},
\end{equation}
then with the spherically symmetric interval
\begin{equation}\label{eq:7}
    ds^2 = -e^{2v(t,r)}dt^2 + e^{2f(t,r)} dr^2 + r^2d\Omega_{\mathbb{S}^2}^2
\end{equation}
the proof amounts to a solving the set of constraints (\ref{eq:3}) on the metric functions, since all dependence on the 2-form coefficients (\ref{eq:6}) can be eliminated (see appendix). The result is then
\begin{equation}\label{eq:8}
    ds^2 = -\bigg(1 - \frac{2M}{r} - 2\lambda r^2\bigg)dt^2 + \bigg(1 - \frac{2M}{r} - 2\lambda r^2\bigg)^{-1} dr^2 + r^2d\Omega_{\mathbb{S}^2},
\end{equation}
where $\lambda = 6\Lambda$. In principle the same technique could be applied to an axisymmetric spacetime to yield (presumably) the Kerr-de Sitter solution without having to appeal to the Ernst equation \cite{13}, although this appears difficult at present. A similar argument implies the only flat cosmology 
\begin{equation}
    ds^2 = -dt^2 + a^2(t)(dx^2 + dy^2 + dz^2)
\end{equation}
that is Einstein, is that for which
\begin{equation}
    \bigg(\frac{d a(t)}{dt}\bigg)^2 = a(t) \frac{d^2a(t)}{dt^2},
\end{equation}
or that exponential expansion is required on spatial slices, which is of course the de Sitter universe. And finally it follows immediately that for the pp-wave spacetime
\begin{equation}
    ds^2 = H(u,x,y)du^2 + 2du dv + dx^2 + dy^2,
\end{equation}
the commutator vanishes if and only if $H(u,x,y)$ is a harmonic function in $x, y$.

It is not difficult to see from (\ref{eq:3}) - by exploiting the Weyl-Schouten theorem - that any locally conformally flat Einstein spacetime is maximally symmetric and hence must coincide with (A)de Sitter (since in this instance the scalar curvature is characterized by $\Lambda$, we can use the theorem of Weinberg \cite{10} that any two maximally symmetric spaces sharing the same scalar curvature must be equivalent up to a coordinate transformation). In the context of the above, this is the unique representation in which the Riemann tensor operators as the identity on the 2-forms modulo the cosmological constant,
\begin{equation}\label{eq:12}
    R^{\mu \nu}_{\rho \sigma} = \frac{\Lambda}{3} \mathbb{I}_{\Lambda^2(\mathcal{M})}.
\end{equation}

\section{Constructing the Endomorphism Group on 2-forms}
Since the result (\ref{eq:4}) and the subsequent arguments regarding solutions of the field equations are of an operational nature, and since these operations are endomorphisms, it suggests the presence of an underlying group structure with (2,2)-tensor representations on $\Lambda^2(\mathcal{M})$. The product operation here is given by tensor contraction, for instance
\begin{equation}
    \omega_{\mu \nu} = \frac{1}{2} R^{\alpha \beta}_{\mu \nu} \varepsilon_{\alpha \beta}^{\gamma \delta} \omega_{\gamma \delta}.
\end{equation}
The Hodge dual has the property (\ref{eq:2}) which restricts its product. However, there is no such constraint for the Riemann tensor, allowing one to construct various 'powers' that retain the endomorphism character by virtue of symmetry,
\begin{equation}\label{eq:14}
    (\mathcal{R})^n \equiv \big(R^{\rho\sigma}_{\mu \nu}\big)^n = \underbrace{R^{ab}_{\mu \nu} R^{cd}_{ab}R^{ef}_{cd} \cdots R^{\rho \sigma}_{\alpha \beta}}_{n\text{-times}}
\end{equation}
We therefore have a well defined identity element (\ref{eq:2}), a product operation (\ref{eq:14}) and associativity which is a direct consequence of index contraction. In order to construct a group, it is necessary that each element have a unique inverse. In the case of the Hodge dual, it is just minus itself by virtue of (\ref{eq:2}). For the powers of the Riemann tensor (and every concomitant with the Hodge dual, which is restricted in Einstein spacetimes by the vanishing commutator), an inverse is defined by the condition
\begin{equation}
    \frac{1}{4}\Phi^{\alpha \beta}_{\mu \nu} R^{\rho \sigma}_{\alpha \beta} = \mathbb{I}_{\mu \nu}^{\rho \sigma}.
\end{equation}
The factor of $1/4$ is necessary due to the symmetry properties of each tensor which could be absorbed into the definition of each object as an operator on 2-forms, as with the Hodge dual (\ref{eq:2}). While we will not so show that such tensors exist in general, for the spacetimes that we will consider (de Sitter, Schwarzschild, Schwarzschild-de Sitter, etc) the inverse $\Phi$ turns out to be the Riemann tensor with all individual elements inverted. All powers have inverse defined in an identical way,
\begin{equation}\label{eq:16}
     \big(\varphi\big)^n \equiv \big(\Phi^{\rho\sigma}_{\mu \nu}\big)^n  = \underbrace{\Phi^{ab \mu \nu}_{ab} \Phi^{cd}_{ab}\Phi^{ef}_{cd} \cdots \Phi^{\rho \sigma}_{\alpha \beta}}_{n\text{-times}}, \quad
    \frac{1}{4^n} \big(\Phi^{\alpha \beta}_{\mu \nu}\big)^n \big(R^{\rho \sigma}_{\alpha \beta}\big)^n  = \mathbb{I}_{\mu \nu}^{\rho \sigma}
\end{equation}
We will refer to both (\ref{eq:14}) and (\ref{eq:16}) generically as 'the Riemann tensors' in what follows. There is an important point here that is worth addressing. There may exist many more tensors that are able to operate as endomorphisms 2-forms. Indeed, the necessary condition is that such a (2,2)-tensor be antisymmetric after contraction with a 2-form in the remaining indices. Here we are interested in operations that are concomitant to the metric and its first and second derivatives, for reasons that will become apparent. While there is no proof (that we know of) that the only such operations to consider are the Riemann tensor together with the Hodge dual (analogous to the Lovelock theorem), we will limit ourselves to these here. 

Bringing everything together, we have a group composed of elements 
\begin{equation}\label{eq:17}
    G = \{ \varphi^n, \mathcal{R}^n, \hodge, \mathbb{I}_{\Lambda^2(\mathcal{M})}\}, 
\end{equation}
such that each element acts as an endomorphism on 2-forms
\begin{equation}
    \forall q \in G, \; q\colon \Lambda^2(\mathcal{M}) \to \Lambda^2(\mathcal{M}),
\end{equation}
with the group multiplication law given by tensor contraction
\begin{equation}
    q \cdot h \mapsto k_{\mu \nu}^{\rho \sigma} \equiv q_{\mu \nu}^{\alpha \beta} h_{\alpha \beta}^{\rho \sigma}.
\end{equation}
Notice that in the limit in which $\mathcal{M}$ is Einstein - by virtue of (\ref{eq:4}) - this group is Abelian. It is at present unclear exactly what this group operation preserves in relation to $\Lambda^2(\mathcal{M})$: one can verify that the inner-product on 2-forms is \textit{not} itself an invariant.

\subsection{Properties of G and Some Examples}
It is instructive at this point to consider the structure of $G$ in some simple spacetimes, in order to understand its general properties. Selecting a spacetime amounts to selecting a representation for $G$ (i.e. a choice of metric), which is unspecified otherwise. The Hodge dual and the Riemann tensors operate on the space of 2-forms in fundamentally different ways: given some specific component of a differential form, the Hodge dual is able to transfer this to its orthogonal co-element (modulo orientation). The Riemann tensors on the other hand are bound (at least for the Einstein spacetimes which we have considered). In other words, the Hodge dual may interchange matrix elements in (\ref{eq:6}) - in a precise way - while the Riemann tensors operate element wise, e.g. 
\begin{equation}
    \begin{split}
        \hodge\colon  & \; \omega_{tr} \to \omega_{\theta \phi} \quad \text{(and vice-versa, modulo multiplicative factors)}, \\
        \big(\mathcal{R}, \varphi\big)^n \colon & \; \omega_{tr} \to \omega_{tr} \quad \text{(modulo multiplicative factors)}.
    \end{split}
\end{equation}
The Hodge dual is the more dynamic operation in this sense, although note that no operations within $G$ ever involve derivatives of the 2-form coefficient functions - which is why one can prove, using (\ref{eq:4}), the result (\ref{eq:8}). The powers of the Riemann tensors therefore always operate in such a way so as to scale these coefficient functions element wise. In addition, each subsequent contraction of the Riemann tensors initiates an overall sign flip as an operation on 2-forms, due to asymmetry over contraction.


As mentioned already, the only instance in which $G$ is immediately known is for spacetimes that are both Einstein and locally conformally flat. These are maximally symmetric and hence equivalent to the de Sitter solution. These symmetry restrictions provide a representation in which the Riemann tensor is proportional to the identity (\ref{eq:12}) due to its Lorentz decomposition (\ref{eq:3}).  It follows immediately that we may write
\begin{equation}\label{eq:21}
    \big(\mathcal{R}\big)^n = \frac{\Lambda^n}{3^n}\mathbb{I}_{\Lambda^2(\mathcal{M})}, \quad \big(\varphi\big)^n = \frac{3^n}{\Lambda^n}\mathbb{I}_{\Lambda^2(\mathcal{M})}.
\end{equation}
Thus, the Riemann tensor acts as a scaling transformation on differential 2-forms, with the characteristic length scale provided by the non-vanishing cosmological constant. It has been argued previously that de Sitter is the unique kinematical ground state of GR, since it is known prior to the solving of the field equations as a moduli space \cite{33}. Viewed as operations, the construction of (\ref{eq:21}) supports this as it is the unique representation in which the cosmological constant completely characterizes $G$; in Minkowski spacetime, the only operation that remains is the Hodge dual, which is trivial in that it induces only sign changes. 

As another example, let us consider those four-dimensional spacetimes that are both Einstein and spherically symmetric. As we have seen, there is a correspondence in this case to the Schwarzschild-(A)de Sitter solution (\ref{eq:8}). Unlike for (\ref{eq:21}) however the scaling is not constant and depends on the 2-form element under consideration. Suppose we wish to compute
\begin{equation}
    \Tilde{\omega}_{t\phi} = R^{\rho \sigma}_{t \phi} \omega_{\rho \sigma}.
\end{equation}
The only non-vanishing component of the Riemann tensor in this instance (modulo asymmetry) is 
\begin{equation}
    R^{t \phi}_{t\phi} = \frac{2\lambda r^3 - M}{r^3}.
\end{equation}
Using (\ref{eq:6}), we again see element-wise scaling,
\begin{equation}
    (\Tilde{\omega})_{t\phi} =  \frac{2\big(2\lambda r^3 - M\big)D(t,r,\theta,\phi)}{r^3} dt \wedge d\phi.
\end{equation}
For comparison, note that 
\begin{equation}
    \big(\star \omega\big)_{t\phi} = \frac{1}{2} \varepsilon_{t\phi}^{\rho \sigma} \omega_{\rho \sigma} = \frac{\big(2\lambda r^3 +2M - r\big) \sin(\theta) C(t,r,\theta,\phi)}{r} dt \wedge d\phi, 
\end{equation}
where we see explicitly the interchanging property of the Hodge dual (functions $D$ and $C$ are permuted). Returning to the Riemann tensor, we may write the general form
\begin{equation}
    \big(R^n\big)^{\rho\sigma}_{t\phi} \omega_{\rho \sigma} = (-1)^{n+1} \Gamma^n D(t,r,\theta, \phi)
\end{equation}
where
\begin{equation}
    \Gamma^n = \bigg[\frac{2 (2\lambda r^3 - M)}{r^{3}}\bigg]^n.
\end{equation}
Note that the factor $2$ is generated by asymmetry. Similar results hold for the remaining components of $\omega$. The inverse powers of the Riemann tensors simply have the inverse scaling relations. When $\Lambda = 0$, we recover the appropriate description of $G$ for the Schwarzschild spacetime. Again, one observes generic scaling behavior: this is indicative of the presence of conformal transformations present in four-dimensional Einstein spacetimes. Indeed, while the group $G$ is interesting in itself, it is actually the contraction over this group that proves to have more surprising properties. In particular, we will find a representation of conformal transformations.

\section{Conformal Transformations as Trace Over G}
The elements of the group $G$ operate as endomorphisms via index contraction on the space of 2-forms. However, as (2,2)-tensor fields there are other permissible operations that one can consider. The tensor trace here plays the essential role. First of all, the Riemann tensor is an element of $G$ and if $\mathcal{M}$ is Einstein its contraction is immediately provided by the Einstein condition
\begin{equation}\label{eq:28}
    \Tr \big(R^{\alpha \nu}_{\mu \beta}\big) \equiv R^{\alpha \nu}_{\mu \alpha}= \Lambda \delta^{\nu}_{\mu}.
\end{equation}
A second point is that the Hodge dual (like the Weyl tensor) has vanishing trace and therefore constitutes the kernel of $G$ under index contraction, together with any element concomitant to it. The remaining elements to consider are therefore the various Riemann tensors and the identity. From (\ref{eq:2}), we can see that the identity is mapped into the Kronecker delta up to a constant factor:
\begin{equation}\label{eq:29}
    \Tr \big(\mathbb{I}_{\Lambda^2(\mathcal{M})}\big) = -\frac{3}{2} \delta^{\nu}_{\mu},
\end{equation}
suggesting that this will play the role of an appropriate identity element of whatever structures remain upon the trace.

Consider now $\mathcal{R}^2$. Two examples of this trace are the following 
\begin{equation}\label{eq:30}
    \Tr \big(\mathcal{R}^2\big) = \begin{cases}
        -\frac{2}{3}\Lambda^2 \delta^{\nu}_{\mu}, \quad & \text{dS} \\
        -\bigg(\frac{M^2}{r^6} + 24 \lambda^2\bigg) \delta^{\nu}_{\mu}, \quad & \text{S-dS}.
    \end{cases}
\end{equation}
In both cases, we can again see proportionality to the Kronecker-delta as with (\ref{eq:28}). The same behaviour is seen in Kerr and Kerr-de Sitter. That is, one has
\begin{equation}\label{eq:31}
    \Tr\big(\mathcal{R}^2\big) = f(x, \Lambda) \delta^{\nu}_{\mu}
\end{equation}
for some function of $f$ of the spacetime coordinates $x^{\mu}$ and the cosmological constant. The same holds for the inverse $\Tr\big(\varphi\big)$ and $\Tr\big(\varphi^2\big)$ although the function is different and not necessarily the inverse of $f$, e.g. 
\begin{equation}
    \Tr \big(\varphi \big) = \begin{cases}
        \frac{9}{\Lambda} \delta^{\nu}_{\mu}, \quad & \text{dS} \\
        \frac{3(2\lambda r^6 + mr^3)}{2(2\lambda^2 r^6 +\lambda mr^3 - m^2)} \delta^{\nu}_{\mu}, \quad & \text{S-dS}.
    \end{cases}
\end{equation}
and more generally we observe
\begin{equation}
    \Tr\big(\varphi^2\big) = \Tilde{g}(x, \Lambda) \delta^{\nu}_{\mu}
\end{equation}
The factors proportional to the cosmological constant in (\ref{eq:30}) come from the following argument. Taking the trace over the second power in the Riemann tensor amounts to computing
\begin{equation}
    \Tr \big(\mathcal{R}^2\big) = R^{\alpha \beta}_{\mu \kappa} R^{\kappa \nu}_{\alpha \beta}.
\end{equation}
Using the Lorentz decomposition of the Riemann tensor (\ref{eq:3}) together with the Einstein condition, we can recast the above in terms of the Weyl tensor,
\begin{equation}\label{eq:36}
    \Tr \big(\mathcal{R}^2\big) = C^{\alpha \beta}_{\mu \kappa} C^{\kappa \nu}_{\alpha \beta} - \frac{2}{3}\Lambda^2 \delta^{\nu}_{\mu}.
\end{equation}
This procedure can be iterated for any power in the curvature tensor. The property (\ref{eq:31}) is, of course, general: it holds to any power of the Riemann tensor and its inverse. That is, for Einstein spacetimes, the trace over any power of the Riemann tensors is proportional to the Kronecker-delta:
\begin{equation}
    \begin{split}
        \Tr\big(\mathcal{R}^n\big) = f^n(x, \Lambda) \delta^{\nu}_{\mu}, \\
        \Tr\big(\varphi^n\big) = \Tilde{g}^n(x, \Lambda) \delta^{\nu}_{\mu},
    \end{split}
\end{equation}
where $f^n, \Tilde{g}^n$ are unique for each choice of $n$. We present a brief argument in the next subsection. For now, one notes that - being proportional to the Kronecker-delta - these traces have the remarkable property that (viewed as an operation) they preserve the metric tensor modulo a function. Taking $\mathcal{R}^n$ for clarity, we have
\begin{equation}\label{eq:37}
    \big[\Tr\big(\mathcal{R}^n\big)\big]^{\nu}_{\mu} g_{\alpha \nu} = f^n(x, \Lambda) g_{\alpha \mu}.
\end{equation}
Provided that we modulo over the sign of these functions (to remove any negative such as that appearing in (\ref{eq:29})) then this is nothing more than a tensor representation of a conformal transformation \cite{15, 17, 18, 19}
\begin{equation}\label{eq:38}
    g_{\mu \nu} \mapsto g'_{\mu \nu} = \Omega^2(x) g_{\mu \nu}.
\end{equation}
Therefore each non-vanishing trace over $G$ (viewed as a contraction operation performed element-wise) generates a conformal transformation of the metric. Each of these conformal transformations can be collected together to again form an Abelian group, since for each transformation (\ref{eq:37}) a unique inverse can be defined relative to the identity as before, which here however is just the Kronecker delta itself (defined in terms of the metric as $g_{\mu \nu}g^{\nu \rho} = \delta_{\mu}^{\rho}$). Consequently, simply by virtue of being curved, a four-dimensional Einstein spacetime has access to (some part thereof) its own conformal equivalence class $[g]$. 

\subsection{The Kronecker-delta Property of the Trace}
That each trace of the elements of $G$ be either 0 or a representation of a conformal transformation is a remarkable property of the Einstein spacetimes that needs justification. Clearly this is an algebraic result unique to the Einstein condition $R_{\mu \nu} = \Lambda g_{\mu \nu}$. Its proof is a consequence of the duality structure of the four-dimensional Lorentzian spacetimes and is a precursor to the Petrov classification. Indeed, Petrov proves \cite{a_1, a_3} that the self-dual part of the Riemann tensor must be an algebraic function of the Ricci tensor. This self-dual component may be written as 
\begin{equation}\label{eq:39}
    \frac{1}{2}\bigg(R_{\mu \nu \rho \sigma} + \frac{1}{4} \varepsilon_{\mu \nu a b} R^{ab cd} \varepsilon_{\rho \sigma c d}\bigg) \equiv R + \star R \star. 
\end{equation}
The vanishing of this self-dual component of the Riemann tensor is then entirely equivalent to the statement of (\ref{eq:4}), which is indeed the statement that a four-dimensional spacetime be Einstein. However, one has to be careful here because we may further decompose the Riemann tensor into the eingenbases of the Hodge dual \cite{9, 15}. In general, it is a symmetric matrix over the product space $\Lambda^2(\mathcal{M}) \otimes \Lambda^2(\mathcal{M})$, decomposable as\footnote{Or the equivalent basis of the bi-vectors. Since they are isomorphic, the choice is irrelevant.}
\begin{equation}\label{eq:40}
    R = \begin{pmatrix}
        A & B \\ 
        B^{T} & C
    \end{pmatrix},
\end{equation}
where A is the self-dual self-dual component, B is the self-dual anti-self-dual component (with $B^T$ its transpose) and C the anti-self-dual anti-self-dual component. In this notation, (\ref{eq:39}) is the the first row of the block decomposition. The Einstein condition (\ref{eq:4}) is equivalent to the vanishing of $B$, the self-dual anti-self-dual component. In addition, $C$ is the complex conjugate of $A$ \cite{9, 15}. Thus, for an Einstein spacetime, all information regarding curvature must be contained in the self-dual self-dual component A of the block decomposition\footnote{It then follows that a spacetime is Einstein if and only if the Levi-Civita connection is self-dual \cite{b_1}.} (\ref{eq:40}), which by Petrov's result must be an algebraic function of the Ricci tensor. The Einstein condition then demands this in turn be proportional to the metric. It follows by raising an index we must have proportionality to the delta function. Furthermore, since the Riemann tensors are contractions over various powers of the Riemann tensor (or its inverse), these too must be algebraic in the Kronecker-delta, and the observation (\ref{eq:36}) holds true: traces over the endomorphism group $G$ on 2-forms yield conformal transformations.

\section{Invariants as Second-Order Contractions over G}
We emphasised in the previous section that viewing the elements of $G$ as tensor fields allows for the standard operations of tensor calculus to be performed element wise over $G$. Initially, elements of $G$ are the $(2,2)$ tensor fields given by (\ref{eq:17}). Taking the contraction (or tensor trace) is a map $(2,2) \to (1,1)$, where for the spacetimes we considered such $(1,1)$-tensor fields are seen to be proportional to the Kronecker delta, and hence form a tensor representation of conformal transformations (\ref{eq:37}). Taken together, all these operations again form a group $G'$ which we may manipulate in the same way. That is, we have reductions
\begin{equation}\label{eq:41}
    A^{\mu \nu}_{\rho \sigma} \in G \; \mapsto \; A^{\nu}_{\rho} \propto \delta^{\nu}_{\rho} \in G' \; \mapsto \; A \in C^{\infty}(\mathbb{R}),
\end{equation}
where each map is the trace defined in (\ref{eq:28}).

The reduction to scalar fields is of interest in both the general and Einstein case. Denoting generically in (\ref{eq:41}) the reduction to scalar fields over the double trace as $\Tr^2(\cdot)$, we draw attention to three specific instances:
\begin{equation}\label{eq:42}
    \begin{split}
        (i)\colon \quad & \Tr^2(R^2) = R^{\mu \nu}_{\rho \sigma} R^{\sigma \rho}_{\mu \nu}, \\
        (ii)\colon \quad & \Tr^2(R \star R) = \frac{1}{2} R^{\mu \nu}_{\rho \sigma} \varepsilon_{\mu \nu}^{\alpha \beta} R_{\alpha \beta}^{\sigma \rho}, \\
        (iii)\colon \quad & \Tr^2(\star R \star R) = \frac{1}{4} \varepsilon^{\mu \nu}_{\rho \sigma} R_{\mu \nu}^{\alpha \beta} \varepsilon_{\alpha \beta}^{\gamma \delta} R^{\sigma \rho}_{\gamma \delta},
    \end{split}
\end{equation}
which constitute the second-order curvature invariants on $\mathcal{M}$ \cite{c_1}. (i) is obviously the Kretschmann scalar, while (ii) and (iii) are less obvious presented in this form. (ii) may be identified (through the Chern-Weil theorem) with the integrand of the first Chern-Pontryagin class, $p_1(\mathcal{M})$ \cite{15, b_1, b_2}. In words, one expects that integration of (ii) over the whole manifold yields exactly an integer, a topological invariant that characterises $\mathcal{M}$. (iii) is also such an integrand, this time the topological Euler density, which via the Gauss-Bonnet theorem we may express as 
\begin{equation}\label{eq:43}
    R^{\mu \nu \rho \sigma}R_{\mu \nu \rho \sigma} - 4R^{\mu \nu}R_{\mu \nu} + R^2 = \Tr^2(\star R \star R).
\end{equation}
For the Einstein spacetimes, something very interesting happens. From (\ref{eq:43}), we first notice that the contribution to the Euler density $R^2 - 4R^{\mu \nu}R_{\mu \nu}$ is identically zero. Alternatively, we may notice that since for Einstein spacetimes the Riemann tensor commutes with the Hodge dual (\ref{eq:4}), the right hand side is simply $\Tr^2(-R^2)$. Therefore, one may define an Einstein spacetime to be exactly those spacetimes whose topological Euler density is completely characterized by its Kretschmann scalar. Additionally, any higher-order curvature invariant constructed in this way can contain at most one power of the Hodge dual. Analyzing the nature of $G$ under trace operations therefore gives a compact and self-consistent approach to construction of (non-derivative) curvature invariants on $\mathcal{M}$. Such analysis could lead to connections between the group $G$ and de-Rham cohomology in the context of the Chern-Weil theorem.

\subsection{Topological Properties of the Hawking Temperature}
That the Kretschmann scalar and topological Euler density are equivalent for Einstein spacetimes has profound implications. Recently Altas and Tekin \cite{d_1} introduced a method for computation of the Hawking temperature that does not rely on semi-classical analysis. One need only construct an integral of the Kretschmann scalar over a temporal slicing of a spacetime region bounded by horizons. The nature of this result is, of course, \textit{topological} (as observed previously, see \cite{d_2, d_4, d_5}) since the equivalence (\ref{eq:43}) demands that the Hawking temperature of a horizon be a global property of an Einstein manifold. One may combine these observations into a simple 'trick' that allows for the computation of either the Euler characteristic $\chi(\mathcal{M})$ or the Hawking temperature of an Einstein spacetime, whichever is not known. The Euler characteristic may be written as 
\begin{equation}
    \chi(\mathcal{M}) = -\frac{1}{128\pi^2} \int_{\mathcal{M}} \Tr^2(\star R \star R) \sqrt{-g} d^4 x,
\end{equation}
or equivalently 
\begin{equation}
    \chi(\mathcal{M}) = \frac{1}{32\pi^2} \int_{\mathcal{M}}  R^{\mu \nu}_{\rho \sigma} R^{\rho \sigma}_{\mu \nu} \sqrt{-g} d^4x.
\end{equation}
We may decompose the measure over the four coordinates, together with the appropriate ranges. For the radial integral, the bounds are determined by the horizons. The only difficult part here is the temporal integral which has infinite bounds (Altas and Tekin remove this trivial divergence by taking a compact time interval which they let diverge at the end of the calculation). The trick - and what all the analysis of previous comments amounts to (formalized in the context of the Wick rotation and Euclidean path integrals) \cite{d_2, d_5} - is that one can simply \textit{replace} the temporal integral with the inverse Hawking temperature. That is, we have
\begin{equation}\label{eq:46}
    \chi(M) = \frac{1}{32\pi^2 T_H} \int^{r_{\Lambda}}_{r_h} dr \int R^{\mu \nu}_{\rho \sigma}R^{\rho \sigma}_{\mu \nu} \sqrt{-g} d\Omega^2.
\end{equation}
Here $r_{\Lambda}$ and $r_h$ set the scale for horizons (if present) generated by a cosmological constant $\Lambda$ and black hole mass $M$, respectively: effectively, they define the region over which Schwarzschild coordinates (or some equivalent exterior cover) are well defined. One can verify this formula for standard cases (e.g. Schwarzschild, S-dS etc) either as a means of computing $\chi(M)$ or $T_H$. Regardless, the key observation is that for the Einstein spacetimes the Hawking temperature of a horizon is of a \textit{topological nature}, independent of any semi-classical quantum field theory one could define over $\mathcal{M}$.

\section{Conclusions}
The four-dimensional Einstein spacetimes derive many of their unique properties from the non-simple structure of the Lorentz algebra $\mathfrak{so}(1,3)$ \cite{9}. Both the Hodge dual and the Riemann tensor act as endomorphisms on the space of 2-forms, with the special limit of their commutator vanishing being equivalent to the Einstein condition $R_{\mu \nu} = \Lambda g_{\mu \nu}$. This allows for a simple generalization of the Birkhoff theorem for spacetimes that are asymptotically (A)dS, giving the Schwarzschild-de Sitter solution (\ref{eq:8}), without appealing to the Einstein field equations. As operations, both the Riemann tensor and the Hodge dual each have a unique inverse since an identity operation is provided on $\Lambda^2(\mathcal{M})$ by (\ref{eq:2}). The Riemann tensor and its inverse can be taken to arbitrary powers via tensor contraction which act as scaling transformations on the differential 2-forms. These operations combine to form an Abelian group $G$ (\ref{eq:17}), which we reiterate has no clear interpretation (at present) as a symmetry of spacetime.

One expects that this group should contain conformal data for the following reason: the Hodge dual on 2-forms determines the conformal metric \cite{9, 30}. In the Einstein limit the group however is Abelian, which suggests that the Riemann tensor should also determine the conformal metric since it commutes as an operation with the Hodge dual and must therefore share 'spectral data'. We have shown that this is the case through the trace operation, which is provided in the first instance by the Einstein condition itself (28). The trace over any element of $G$ that is concomitant to the Hodge dual vanishes, while on the other hand we have argued (thanks to a result of Petrov \cite{a_1}) that the non-vanishing elements are always proportional to the Kronecker-Delta (\ref{eq:36}) for the Einstein spacetimes. More fundamentally, the reason this result holds is that these traces contain the conformal data: since they are proportional to the Kronecker-Delta, contraction with the metric tensor preserves the metric up to some functional multiplier (\ref{eq:37}), which is precisely a tensor representation of the conformal transformations \cite{15, 17, 18, 19}. Since the trace over $G$ removes the Hodge dual, the remaining objects are required to determine the conformal metric by the arguments above, which can only be done via this proportionality since anything else would spoil the property (\ref{eq:37}). Note that the collection of all such elements $\Tr(G)$ again forms an Abelian group, since the identity is provided by the Kronecker delta (\ref{eq:29}). We are lead therefore to conclude that the Einstein spacetimes have access to (somepart thereof) their own conformal equivalence class simply by virtue of being curved. The de Sitter solution forms the ground state of GR not only in the kinematic sense \cite{30}, but also here as the fundamental representation (\ref{eq:12}) known via maximal symmetry. In this case, upon taking the trace each element of $G$ resolves to a scaling transformation (homothety) in powers of the cosmological constant, showing its active role as a generator of such transformations. The "somepart thereof" in our initial statement is important since not every conformal transformation is necessarily generated in the manner devised here, as is clear from the de Sitter example, but some natural subset are nonetheless present. However, we recall that in the construction of $G$ we considered only the Hodge dual and Riemann tensors: it is possible that there are other tensors (not concomitant to the metric) that act as endomorphisms of $\Lambda^2(\mathcal{M})$, which when contracted in the manner described populate the conformal group in its entirety, but this is a subject requiring further investigations.

Tracing again over the group yields characteristic curvature invariants of $\mathcal{M}$, which by some important theorems are - at the quadratic level (\ref{eq:42}) \cite{c_1} - the associated integrands of topological invariants, notably the first Chern-Pontryagin class, $p_1(\mathcal{M})$ and the Euler characteristic $\chi(\mathcal{M})$ \cite{b_1, b_2}. Since the Einstein condition is equivalent to the Abelian nature of $G$, for such spacetimes the curvature invariants can contain at most one power of the Hodge dual, with its actual position of contraction being irrelevant. Notably, the Kretschmann scalar and the topological Euler density coincide for such spacetimes. By virtue of a simple trick hinted at in other authors works\cite{d_1, d_2, d_4, d_5}, one can use this form to construct a relationship (\ref{eq:46}) between the Euler characteristic and the Hawking temperature of $\mathcal{M}$, showing that even in four-dimensions the associated temperature of a horizon is a topological property.

\subsection{Open Questions}
The primary goal of this work was to introduce the group $G$ and draw the communities attention to the relationships it incites between group theory, conformal symmetry and global topology, all within the context of the Einstein spacetimes. Many interesting open questions remain: what is the consequence of an Einstein spacetime having access to its own conformal equivalence class? Are higher curvature invariants analogous to (\ref{eq:42}) generated from $G$ necessarily topological?  Can expression (\ref{eq:46}) be formalised?  And perhaps the most important question demanding answer: what is the physical invariant of the endomorphism group $G$?  We will continue to explore these problems in future works.

\begin{appendices}

\section{Proof of Theorem 1}
To prove the generalization of the Birkhoff theorem to asymptotically (A)de Sitter spacetimes, we exploit the fact that a spacetime is Einstein (in four dimensions) if and only if the Riemann tensor commutes with the Hodge dual on 2-forms. Taking the ansatz (\ref{eq:7}), together withe the general 2-form (\ref{eq:6}), one computes 
\begin{equation}
    \omega_{\mu \nu} = R^{\alpha \beta}_{\mu \nu} \varepsilon^{\rho \sigma}_{\alpha \beta} \omega_{\rho \sigma} - \varepsilon^{\alpha \beta}_{\mu \nu} R^{\rho \sigma}_{\alpha \beta} \omega_{\rho \sigma}.
\end{equation}
Demanding that this vanish gives a set of six constraints that can be solved for the unknown metric functions appearing in (\ref{eq:7}). Actually, only three of these constraints carry unique information and we may solve then the system given by 
\begin{equation}\label{A2}
    2D e^{2f}\frac{\partial f}{\partial t} - E e^{2v} \frac{\partial f}{\partial r} - E e^{2v} \frac{\partial v}{\partial r}  = 0,
\end{equation}
\begin{equation}\label{A3}
    2E \frac{\partial f}{\partial t} - D \frac{\partial f}{\partial r} - D\frac{\partial v}{\partial r}  = 0,
\end{equation}
\begin{dmath}\label{A4}
    F\bigg(r^2 e^{2f}\bigg(\frac{\partial f}{\partial t}\bigg)^2 - r^2 e^{2f} \frac{\partial f}{\partial t} \frac{\partial v}{\partial t} + r^2 e^{2v} \frac{\partial f}{\partial r}\frac{\partial v}{\partial r} - r^2 e^{2v}\bigg(\frac{\partial v}{\partial r}\bigg)^2 e^{-(f+v)} + r^2 e^{2f} \frac{\partial^2 f}{\partial t^2} - r^2 e^{2v} \frac{\partial^2 v}{\partial r^2} - \bigg(e^{2f} - 1\bigg) e^{2v}\bigg) e^{-(f+v)}  = 0.
\end{dmath}
where $D, E$ and $F$ are the 2-form coefficient functions (\ref{eq:7}). We can combine equations (\ref{A2}) and (\ref{A3}) to eliminate this dependence, giving 
\begin{equation}\label{A5}
    \pm e^f \frac{\partial f}{\partial t} - \frac{\partial f}{\partial r} - \frac{\partial v}{\partial r} = 0.
\end{equation}
This alone is not enough to proceed. However, the Einstein condition is equivalent to the vanishing of the trace-free Ricci tensor and hence the conditions ({\ref{A2}-\ref{A4}}) must be complimented by this. For the spacetime (\ref{eq:6}) the temporal-radial components of the trace-free Ricci tensor have the form
\begin{equation}
    \frac{2}{r}\frac{\partial f}{\partial t} dt\otimes dr + \frac{2}{r}\frac{\partial f}{\partial t} dr\otimes dt
\end{equation}
 and thus to vanish we must have that $f$ is a function of $r$ alone
\begin{equation}
    f = f(r).
\end{equation}
With (\ref{A5}) then this condition implies that\footnote{One must of course verify that the branch of solutions given by $v = v(t) + c_2(r)$ has no interesting solutions, which is indeed the case.} 
\begin{equation}
    v = v(r) + c_1(t)
\end{equation}
where $c_1(t)$ is some constant function in time, which we set to $0$ without loss of generality. It follows that 
\begin{equation}\label{A9}
    f(r) = -v(r).
\end{equation}
With this, equation (\ref{A4}) becomes a second-order ordinary differential equation in $v$,
\begin{equation}
    2 r^2 \bigg(\frac{dv(r)}{dr}\bigg)^2 + r^2 \frac{d^2v(r)}{dr^2} + e^{-2v(r)} - 1 = 0,
\end{equation}
with analytic solution of the form
\begin{equation}\label{A11}
    v(r) = \frac{1}{2} \ln\bigg(1 - \frac{2M}{r} - 2\lambda r^2\bigg),
\end{equation}
where $M$ and $\lambda$ are integration constants. Consequently, the metric (\ref{eq:7}) is constructed from (\ref{A11}) yielding 
\begin{equation}
    ds^2 = -\bigg(1 - \frac{2M}{r} - 2\lambda r^2\bigg)dt^2 + \bigg(1 - \frac{2M}{r} - 2\lambda r^2\bigg)^{-1} dr^2 + r^2d\Omega_{\mathbb{S}^2}.
\end{equation}
The integration constants $M$ and $\Lambda = 6\lambda$ are thus to be interpreted as the mass and the cosmological constant, respectively. In the limit where $\Lambda = 0$, we have thus recovered the Schwarzschild spacetime with analysis forming the statement of the Birkhoff theorem, but without relying on the form of the Einstein field equations.
\end{appendices}

\end{document}